\documentclass[english,journal=jctcce,manuscript=article,email=true,etalmode=truncate,maxauthors=0]{achemso}
  \usepackage{geometry} 
  \usepackage{stackengine}
  \usepackage{titlesec}

  \usepackage[T1]{fontenc} 
  \usepackage{amsmath}     
  \usepackage{amssymb}
  \usepackage{bm}
  \usepackage{mathrsfs}
  \usepackage{etoolbox}
  \usepackage{environ}
  \usepackage[version=4]{mhchem}
  \usepackage{booktabs}
  \usepackage{adjustbox}
  \usepackage{hhline}

  \usepackage{setspace}    
  \usepackage{float}       
  \newfloat{chart}{htbp}{loc}
  \newfloat{graph}{htbp}{loh}
  \newfloat{scheme}{htbp}{los}

  \usepackage{graphicx}    

  \usepackage[english]{babel}
  \usepackage{array}

  \usepackage{subcaption}  
  \usepackage{outlines}    
  \usepackage[normalem]{ulem} 

  \usepackage[numbers,sort&compress,super]{natbib}
  \usepackage{natmove}

  \usepackage{cancel}      
  \usepackage{pdfpages}  
  \usepackage[font=scriptsize,labelfont=bf]{caption}

  \usepackage{hyperref} 
  \usepackage{cleveref}	
  	\crefname{figure}{Figure}{Figures}
  	\crefname{table}{Table}{Tables}
  	\crefname{equation}{Eq.}{Eqs.}
  	\crefname{section}{Section}{Sections}
  	\crefname{subsection}{Section}{Sections}
  	\crefname{subsubsection}{Section}{Sections}
  	\crefname{algorithm}{Algorithm}{Algorithms}

  \usepackage{tikz}

    \usepackage[ruled,vlined]{algorithm2e}
  
  \newcommand*{\bigO}[1]{\ensuremath{\mathcal{O}({#1})}}

  \usepackage{todonotes}

\author{Karl Pierce}
  \affiliation{ Department of Chemistry, Virginia Tech, Blacksburg, Virginia 24061, U.S.A. }
  \author{Edward F. Valeev}
  \email{efv@vt.edu}
  \affiliation{ Department of Chemistry, Virginia Tech, Blacksburg, Virginia 24061, U.S.A. }
  \title{Efficient construction of canonical polyadic approximations of tensor networks}

\begin{document}

\maketitle
\begin{abstract}
We consider the problem of constructing a canonical polyadic (CP) decomposition for a tensor {\em network}, rather than a single tensor. We illustrate how it is possible to reduce the complexity of constructing an approximate CP representation of  the network by leveraging its structure in the course of the CP factor optimization.
The utility of this technique is demonstrated for the order-4 Coulomb interaction tensor approximated by 2 order-3 tensors via an approximate generalized square-root (SQ) factorization, such as density fitting or (pivoted) Cholesky. The complexity of constructing a 4-way CP decomposition is reduced from $\mathcal{O}(n^4 R_\text{CP})$ (for the non-approximated Coulomb tensor) to $\mathcal{O}(n^3 R_\text{CP})$ for the SQ-factorized tensor, where $n$ and $R_\text{CP}$ are the basis and CP ranks, respectively. This reduces the cost of constructing the CP approximation of 2-body interaction tensors of relevance to accurate many-body electronic structure by up to 2 orders of magnitude for systems with up to 36 atoms studied here.
The full 4-way CP approximation of the Coulomb interaction tensor is shown to be more accurate than the known approaches utilizing CP-decomposed SQ factors (also obtained at the $\mathcal{O}(n^3 R_\text{CP})$ cost), such as the algebraic pseudospectral and tensor hypercontraction approaches. The CP decomposed SQ factors can also serve as a robust initial guess for the 4-way CP factors.
\end{abstract}

\section{Introduction}
\label{sec:introduction}
The canonical polyadic (CP) decomposition\cite{VRG:hitchcock:1927:JMP,VRG:kiers:2000:JC}, also known as CANDECOMP\cite{VRG:carroll:1970:P}, PARAFAC\cite{VRG:harshman:1970:WPP}, separated representation,\cite{VRG:beylkin:2002:PNAS} and other names\cite{VRG:mocks:1988:ITBE}, is a popular data-sparse representation of higher-order tensors as a data analysis tool in many settings as well as a way to reduce the cost and/or complexity of the tensor algebra in some physical simulation applications.
The CP decomposition of an order-$n$ tensor can be viewed as a {\em tensor network} (TN) of $n$ factor matrices, each describing a single mode of the target tensor.
What makes the CP TN unique is that each factor matrix is connected directly to {\em every} other factor via a single {\em hyperedge}.
By treating all modes of the target tensor on equal footing the CP decomposition has a universal structure that simplifies its use compared to other TNs with more specialized topologies (such as matrix product states, etc.).  
Formally, the CP decomposition reduces the storage complexity of higher-order tensors from exponential to linear in the tensor order. However, finding the minimal extent of the CP hyperedge (CP {\em rank})
sufficient for exact CP representation of a given tensor or its approximate CP representation to a fixed given precision is hard.\cite{Hastad:1990:algorithm, VRG:hillar:2013:JA}

Whether CP decomposition and/or TNs with CP-like hyperedges can be gainfully used in quantum many-body simulation, despite the challenges of computing the CP decomposition, is an open question.
Beylkin and Mohlenkamp motivated broad utility of polyadic separation of multidimensional functions and operators (i.e., an infinite-dimensional analog of CP decomposition)
and illustrated its effectiveness for model quantum simulation problems.\cite{VRG:beylkin:2002:PNAS} The only uses of the CP decomposition for {\em global} approximation of the electronic Hamiltonian and/or wave function have been demonstrated by Auer and co-workers;\cite{VRG:benedikt:2011:JCP,Benedikt:2013:JCP,Bohm:2016:JCP} the scope of these applications was limited to small systems by the high cost of the computing and updating the CP decomposition. More limited uses of the CP decomposition appeared in the context of compressed spectral element representation for constructing numerical (real-space) orbitals by Bischoff and Valeev\cite{VRG:bischoff:2011:JCP} and for compressing real-space representations of various quantities in the electronic structure context by Chinnamsetty et al.\cite{VRG:chinnamsetty:2007:JCP} 
More common scenario in the quantum many-body simulation is the appearance of tensor networks featuring CP-like hyperedges, e.g., as a result of approximating an integral by a quadrature.
Such examples include the approximation of the Coulomb operator in the pseudospectral (PS) family of methods,\cite{VRG:friesner:1985:CPL} the related tensor hypercontraction (THC) factorization of 2-particle tensors\cite{VRG:hohenstein:2012:JCP} and its close cousin, interpolative separable density fitting,\cite{VRG:lu:2015:JCP} robust CP-DF factorization marrying the PS and THC approaches,\cite{VRG:pierce:2021:JCTC} and the CD-SVD compression of the Coulomb interaction tensor\cite{VRG:peng:2017:JCTC}. These decompositions have productive uses in quantum simulation contexts on both classical\cite{VRG:neese:2009:CP,VRG:parrish:2014:JCP,VRG:hu:2017:JCTC,VRG:pierce:2021:JCTC,VRG:peng:2017:JCTC} and quantum hardware.\cite{VRG:goings:2022:PNASU} Lastly, the use of Laplace transform in the electronic structure\cite{VRG:almlof:1991:CPL} also leads to networks with CP-like hyperedges.

Our recent experiences\cite{VRG:pierce:2021:JCTC} suggests that more productive uses of the CP factorization can be found when focusing on approximating not {\em individual} tensors but on tensor {\em networks}. Thus here we consider a problem of constructing a CP approximation for a tensor network. To the best of our knowledge such class of problems has not been yet considered. It is clear that leveraging the structure of the tensor network in the course of constructing its CP approximation can lead to notable savings. Here we show that that is indeed the case in a practically-important scenario; namely, we show how to construct a full (4-way) CP decomposition of a 2-particle interaction represented in a molecular orbital basis with reduced complexity by leveraging the approximate generalized square root (SQ) factorization of such a tensor. This allows us to construct full CP decomposition of 2-particle interaction tensors for systems of unprecedented size.
Such productive combination of CP and other tensor network approximations can be further improved by other CP acceleration strategies such as
parallelization\cite{VRG:karlsson:2016:PC} and improved solver strategies.\cite{VRG:acar:2011:JC,VRG:battaglino:2018:SJMA&A,VRG:ma:2022:NLAA}

The rest of the manuscript is organized as follows. \Cref{sec:formalism} introduces the mathematical formalism for SQ-accelerated CP4 decomposition and its application to the 2-body interaction tensors in quantum electronic structure. Technical details of the computational experiments are described in \cref{sec:compdetails}. In \cref{sec:results}  we compared the accuracy of the CP4 decomposition to the more common 3-way CP decompositions as well as illustrate the efficiency of the accelerated CP4 solver relative to its standard (naive) counterpart. Our findings are summarized in \cref{sec:summary}.

\section{Formalism}
 \label{sec:formalism}
Our objective is to approximate
a 2-particle interaction represented in a generic basis $\{\phi_p\}$:
\begin{align}
\label{eq:coul}
    g_{st}^{pq} = \iint \phi^*_s({\bf r}_1) \phi^*_t({\bf r}_2) g({\bf r}_1, {\bf r}_2) \phi_p({\bf r}_1) \phi_q({\bf r}_2) d{\bf r}_1 d{\bf r}_2;
\end{align}
all numerical experiments in this work will use $g({\bf r}_1,{\bf r}_2) \equiv |{\bf r}_1-{\bf r}_2|^{-1}$ (the Coulomb interaction), but other positive kernels can be used.
A CP factorization of $g$ is represented using a set of factor matrices ${\bf x} =\{\pmb{\gamma},\pmb{\rho},\pmb{\beta},\pmb{\kappa}\}$: 
\begin{align}
\label{eq:g_CP4}
    g^{pq}_{st} \overset{\text{CP4}}{\approx} \hat{g}^{pq}_{st} = \sum_{r}^{R_{\text{CP4}}} \lambda_r \gamma^{p}_{r} \rho^{q}_{r} \beta_{s}^{r} \kappa_{t}^{r};
\end{align}
moniker ``CP4'' will be used to help distinguish such 4-way CP factorization from other factorizations involving CP decomposition.

In this work we impose the CP factors be column-wise unit-normalized, i.e. $\sum_p|\gamma^p_r|^2 = 1$, etc.
In general, the CP ``singular values'' $\lambda_r$ can be
absorbed into the factors;\footnote{It should be noted that though $\lambda_r$s resemble singular values, the CP decomposition does not have the same properties as the singular values decomposition\cite{VRG:kolda:2009:SR}} 
the weighted factors will be denoted
by a {\em caron}, $\check{\gamma}^p_r \equiv \lambda_r \gamma^p_r$.
Thus, \cref{eq:g_CP4} then can be written equivalently using weighted factors:
\begin{align}
g^{pq}_{st} \overset{\text{CP4}}{\approx} \sum_{r}^{R_{\text{CP4}}} \check{\gamma}^{p}_{r} \rho^{q}_{r} \beta_{s}^{r} \kappa_{t}^{r} =
\sum_{r}^{R_{\text{CP4}}} \gamma^{p}_{r} \check{\rho}^{q}_{r} \beta_{s}^{r} \kappa_{t}^{r} =
\sum_{r}^{R_{\text{CP4}}} \gamma^{p}_{r} \rho^{q}_{r} \check{\beta}_{s}^{r} \kappa_{t}^{r} =
\sum_{r}^{R_{\text{CP4}}} \gamma^{p}_{r} \rho^{q}_{r} \beta_{s}^{r} \check{\kappa}_{t}^{r}  
\end{align}

There exists no finite algorithm to determine the exact or approximate (sufficient for a given accuracy) CP rank of a general tensor\cite{Hastad:1990:algorithm,VRG:hillar:2013:JA}. Thus in practice the CP rank is determined heuristically. For a fixed CP rank \cref{eq:g_CP4} is typically constructed by minimizing the norm of the error,
\begin{align}
\label{eq:fCP}
f(\mathbf{x}) \equiv \frac{1}{2} \sum_{pqst} |g^{pq}_{st} - \hat{g}^{pq}_{st}|^2
.
\end{align}
The nonlinear optimization uses well-known heuristics (Alternating Least Squares (ALS)\cite{VRG:carroll:1970:P,Harshman:1970:WPP,VRG:beylkin:2002:PNAS}, accelerated gradient\cite{Acar:2011:JC,Phan:2013:SJMAP,Tichavsky:2013:ICASSP,VRG:benedikt:2011:JCP,Espig:2012:NM}, nonlinear least squares\cite{VRG:sorber:2013:SJO}, etc.) whose cost-determining step is the evaluation of the objective function's gradient (or its components):
\begin{align}
\label{eq:fCP-grad}
    \frac{\partial f}{\partial \check{\gamma}^r_p} = & \, 2 \sum_{qst}\left( \hat{g}_{pq}^{st} - g_{pq}^{st} \right) \rho^{q}_{r} \beta_{s}^{r} \kappa_{t}^{r} 
    = \, 2 \left ( \sum_{r'} \check{\gamma}_{p}^{r'} \left( \mathbf{W}^{\{\gamma\}}\right)_{r'}^{r}  
    - \sum_{qst} g_{pq}^{st} \rho^{q}_{r} \beta_{s}^{r} \kappa_{t}^{r} \right),
\end{align}
where
\begin{align}
    \left( \mathbf{W}^{\{\gamma\}}\right)^{r}_{r'} \equiv \left(\sum_q \rho_{q}^{r'} \rho^{q}_{r}  \right) \left(\sum_s  \beta^{s}_{r'} \beta_{s}^{r} \right) \left(\sum_t \kappa^{t}_{r'} \kappa_{t}^{r}  \right).
\end{align}
The simplest way to update CP factors is via ALS, by (1) solving the linear least squares problem $ \partial f/\partial \check{\gamma}^r_p =0$ for $\check{\gamma}$ \footnote{Since matrix ${\bf W}$ is Hermitian, fast (pivoted) Cholesky solver for the linear system can be used.},
(2) normalizing its columns to obtain updated $\gamma$ and $\lambda$, and repeating for other factors until the desired ALS convergence criterion ($\epsilon_{\rm ALS}$ has been achieved (here we use the criterion of Ref. \cite{VRG:pierce:2021:JCTC}), and repeating as necessary.
Computing the last term in \cref{eq:fCP-grad}, which is needed both in ALS and in the gradient-based approaches, is the cost-determining step, with a complexity of \bigO{N^4 R}, where $N$ is proportional to the mode dimension and $R$ is the CP rank (in practice $R \propto \bigO{N}$).
The high-order cost of the gradient evaluation is aggravated by the difficulty of the optimization problem (strong nonlinearity, sensitivity to the initial guess, and often poor conditioning; e.g., see Ref. \citenum{Navasca:2008:JCP})
These factors explain why the CP4 decomposition has barely been explored in the context of quantum simulation, as mentioned in \cref{sec:introduction}.

Popular CP-like approximations of $g$ can be determined by starting from the generalized\footnote{A square root $B$ of matrix/operator $A$ satisfies $A=BB$.  In some contexts, such as Kalman filtering, $B$ such that $A=B^\dagger B$, for positive $A$, is also referred to as square root. To avoid confusion we will be refer to such $B$ as generalized square root of $A$.} square root (SQ) factorization:
\begin{align}
\label{eq:DFints}
    g^{pq}_{st} \overset{\text{SQ}}{\approx} \sum_{X} B^{pX}_{s} B^{qX}_{t}.
\end{align}
The two most popular ways to determine the square root
factor $B$ are the density fitting \cite{VRG:whitten:1973:JCP,VRG:dunlap:1979:JCP,Vahtras:1993:CPL,VRG:jung:2005:PNAS} (DF) approximation, in which 
$X$ represents a predetermined fitting (``auxiliary'') basis that grows linearly with the system size, and the rank-revealing Cholesky decomposition (CD).\cite{VRG:beebe:1977:IJQC,VRG:lowdin:1965:JMP,Lowdin:2009:IJQC,Folkestad:2019:JCP} 
The error of \cref{eq:DFints} can be controlled in the case of DF by using a more complete fitting basis, or in the case of CD by increasing the Cholesky rank.

Although the generalized square root factorization is extremely popular, due to its ability to reduce the storage complexity of $g$ from $N^4$ to $N^3$, it rarely reduces the computational complexity of electronic structure methods by itself.
However, by combining \cref{eq:DFints} with the CP decomposition it is possible to reduce the computational complexity for several important algorithms, such as the exact exchange evaluation.\cite{VRG:friesner:1985:CPL,Lee:2019:JCTC}
Traditionally, this is accomplished by CP factorizing the tensor $B$:
\begin{align}
\label{eq:SQ}
    B^{pX}_{s} \approx \hat{B}^{pX}_{s} = \sum_r^{R_{\text{CP3}}} \lambda_r \gamma^p_r \beta_s^r \chi^X_r = \sum_r^{R_{\text{CP3}}} \gamma^p_r \beta_s^r \check{\chi}^X_r,
\end{align}
where $R_{\text{CP3}}$ is the CP rank of this decomposition; we will refer to a 3-way CP decomposition by moniker ``CP3''. Note that although \cref{eq:SQ} does not impose the hermiticity of $B$ (with respect to swapping of $p$ and $s$), in practice
$\gamma^p_r = \left(\beta^r_p\right)^*$ can be imposed.\cite{VRG:hummel:2017:JCP,VRG:pierce:2021:JCTC}
The CP3 approximation can then be introduced into \cref{eq:DFints}:
\begin{itemize}
\item {\em once}
\begin{align}
    \label{eq:g_PS}
    g^{pq}_{st} \approx \sum_{X} \hat{B}^{pX}_{s} B^{qX}_{t} 
    = \sum_{r}^{R_{\text{CP3}}}  \gamma^{p}_r \beta_{s}^{r} \left( \sum_X \check{\chi}^X_r B^{qX}_{t} \right) \equiv \sum_{r}^{R_{\text{CP3}}} \gamma^{p}_r \beta_{s}^{r} \tilde{B}^{t}_{qr} ,
\end{align}
to produce a factorization of the pseudospectral (PS) method\cite{VRG:friesner:1985:CPL,Friesner:1986:JCP,Langlois:1990:JCP,VRG:ringnalda:1990:JCP,Friesner:1991:ARPC,Martinez:1995:JCP,VRG:martinez:1992:JCP,Martinez:1994:JCP,Ko:2008:JCP,Martinez:1993:JCP};
\item {\em twice},
\begin{align}
\label{eq:g_THC}
    g^{pq}_{st} \approx \sum_{X} \hat{B}^{pX}_{s} \hat{B}^{qX}_{t} =
    \sum_{r,r'}^{R_{\text{CP3}}} \gamma^{p}_r \beta_s^r \gamma^q_{r'} \beta_t^{r'} \left(\sum_X \check{\chi}^X_{r}  \check{\chi}_{r'}^X\right) \equiv \sum_{r,r'}^{R_{\text{CP3}}} \gamma^{p}_r \beta_s^r \gamma^q_{r'} \beta_t^{r'} \tilde{\chi}_{r,r'} ,
\end{align}
to produce the tensor hypercontraction format\cite{VRG:hohenstein:2012:JCP,VRG:hohenstein:2012:JCPa,VRG:parrish:2012:JCP,Hohenstein:2013:JCP,VRG:parrish:2014:JCP,Shenvi:2013:JCP,Schutski:2017:JCP,Parrish:2019:JCP,Lee:2019:JCTC,VRG:hummel:2017:JCP}; or
\item {\em robustly}\cite{VRG:pierce:2021:JCTC} using a combination of single and double substitutions,
\begin{align}
\label{eq:g_rCP3DF}
    g^{pq}_{st} \approx \sum_{X} \hat{B}^{pX}_{s} B^{qX}_{t} + \sum_{X} B^{pX}_{s} \hat{B}^{qX}_{t} - \sum_{X}\hat{B}^{pX}_{s} \hat{B}^{qX}_{t}.
\end{align}
\end{itemize}
The latter, denoted as rCP3-DF, has been shown recently\cite{VRG:pierce:2021:JCTC} to be more accurate than either \cref{eq:g_PS} or \cref{eq:g_THC}.
As we demonstrate next, the ``square-root'' factorization can be also useful as a {\em precursor} to the CP4 approximation of $g$. Specifically, its use allows one to reduce the complexity of constructing the CP4 approximation from \bigO{N^5} to \bigO{N^4}.

As mentioned above, evaluation of the gradient of the CP objective function, \cref{eq:fCP-grad}, costs \bigO{N^4 R}.
However, it is possible to reduce the complexity of the CP optimization problem by replacing tensor $g$ with its SQ-factorization, \cref{eq:DFints}:
\begin{align}
\label{eq:fCP-grad-SQ}
\frac{\partial f}{\partial \check{\gamma}^r_p} \overset{\rm SQ}{=} & \, 2 \sum_{qst}\left( \hat{g}_{pq}^{st} - \sum_X \sum_{X} B^{s}_{pX} B^{t}_{qX} \right) \rho^{q}_{r} \beta_{s}^{r} \kappa_{t}^{r} \nonumber \\
    = & \, 2 \left ( \sum_{r'} \check{\gamma}_{p}^{r'} \left( \mathbf{W}^{\{\gamma\}}\right)_{r'}^{r}  
    - \sum_X \left( B^{s}_{pX} \beta_{s}^{r} \right) \left( \sum_{qt} B_{qX}^{t}  \rho^{q}_{r} \kappa_{t}^{r} \right)  \right).
\end{align}
Using the SQ-network approximated CP4 loss function, we reduce the cost of the  gradient of CP4 from \bigO{N^4R} to \bigO{N^2 R X}. 
Furthermore, we reduce the storage requirements of the approach from \bigO{N^4 \approx N^3 R} to \bigO{N^3 \approx N^2 R}
Although others have introduced TNs into the CP loss function\cite{Schutski:2017:JCP}, this work is first to construct factor matrices for {\em only} the external modes of a TN.
The CP decomposition, in this way, provides a route to approximate the result of a TN without \textit{a priori} construction of the tensor from the network.
Clearly, such use of CP can be more generally useful than just for the case of the Coulomb integrals.

Now, formally the CP4 ALS optimization has the same computation scaling as the CP3 ALS optimization, \bigO{N^4}.
However, it is known that the efficiency of the CP optimization process and the accuracy of the resulting factorization can strongly depend on an initial guess.
In this work we utilize an initial guess scheme which uses the CP3-SQ-factorization of $g$ as well as vectors filled with quasi-random numbers taken from a uniform distribution on [-1,1].
To motivate our initialization strategy, we compare the CP4-approximated $g$ to its THC-like CP3-SQ counterpart:
\begin{align}
\label{eq:g_CP4_1}
g^{pq}_{st} \overset{\rm \cref{eq:g_CP4}}{\approx} & \sum_{r}^{R_{\text{CP4}}} \gamma^{p}_{r} \beta_{s}^{r} \rho^{q}_{r} \kappa_{t}^{r} \lambda_r \\
\label{eq:g_THC_1}
g^{pq}_{st} \overset{\rm \cref{eq:g_THC}}{\approx} & \sum_{r,r'}^{R_{\text{CP3}}} \gamma^{p}_r \beta_s^r \gamma^q_{r'} \beta_t^{r'} \tilde{\chi}_{r,r'}
\end{align}
The key differences between these expressions are: (1) factors $\gamma$ and $\beta$ in \cref{eq:g_CP4_1,eq:g_THC_1} were determined to CP-approximate different tensors ($g^{pq}_{st}$ and $B^{pX}_s$, respectively), and (2) \cref{eq:g_THC_1} is explicitly particle-symmetric. Importantly, however, the CP3-SQ approximation (\cref{eq:g_THC_1}) can be rewritten in the CP4 form {\em exactly} by merging indices into a composite index $w\equiv\{r,r'\}$:
\begin{align}
\label{eq:g_THC_as_CP4}
    g^{pq}_{st} \overset{\rm \cref{eq:g_THC_1}}{\approx} & \sum_{w}^{R^2_{\text{CP3}}} \gamma^{p}_w \beta_s^w \rho^q_{w} \kappa_t^{w} \lambda_{w}
\end{align}
where
\begin{align}
\gamma^{p}_w \equiv& \gamma^{p}_{\{r,r'\}} \equiv \gamma^{p}_r \otimes \mathbb{I}_{r'}, \\
\beta_{s}^w \equiv& \beta_s^{\{r,r'\}} \equiv \beta_s^r \otimes \mathbb{I}_{r'}, \\
\rho^{q}_w \equiv& \rho^{q}_{\{r,r'\}} \equiv \gamma^{q}_{r'} \otimes \mathbb{I}_{r}, \\
\kappa_{t}^w \equiv& \kappa_t^{\{r,r'\}} \equiv \beta_t^{r'} \otimes \mathbb{I}_{r}. \\
\lambda_{w} \equiv& \tilde{\chi}_{r,r'}
\end{align}
such that $\otimes$ represents the outer product and $\mathbb{I}_{r}$ is an identity vector with dimension $r$.
In practical applications of THC the CP3 rank is \bigO{N}, thus it may appear that the CP4 rank required to represent it exactly is \bigO{N^2}. However since FMM\cite{VRG:greengard:1987:JCP} and related approximations\cite{VRG:beylkin:2008:ACHA} can be used to construct representations of sufficiently-well-behaved integral operators with \bigO{N \log{N}} complexity, this limits the asymptotic CP4 rank of $g$ to \bigO{N \log N} as well. 
In the context of THC, this asymptotic limit means that the $\tilde{\chi}$ matrix can be made sparse for sufficiently large systems by an appropriate transformation. 
However, for small systems where the $\tilde{\chi}$ matrix is dense,
direct construction of CP4 can be more compact than the THC representation, i.e. a CP4 approximation with an accuracy of $\epsilon_\mathrm{THC}$ can be found via an optimization of a rank $R_\mathrm{CP4} << R_\mathrm{CP3}^2$ CP4-ALS decomposition (see \cref{sec:results} for details).

The formal connection between \cref{eq:g_CP4} and \cref{eq:g_THC_1} via \cref{eq:g_THC_as_CP4} thus suggests the use of unit-normalized CP3 factors as the initial guess for the CP4 decomposition.
Further, because the CP optimization recomputes $\lambda$ every iteration, it is not necessary to keep elements of the diagonalized $\tilde{\chi}$ matrix (in \cref{eq:g_THC_1}).
If the target CP4 rank is greater than the CP3 rank,
the CP4 factors can be padded with vectors of quasi-random numbers taken from the uniform distribution on [-1,1] and normalized to unity.
This is sufficient to bootstrap the ALS CP4 solver since, again, the singular values are not needed to compute the gradient with respect to the weighted factors (see \cref{eq:fCP-grad}).

\section{\label{sec:compdetails} Computational details}
The CP3 and CP4 optimization computations have been computed using the Basic Tensor Algebra Subroutine (BTAS) software package\cite{BTAS}.
All of these computations were run on the Virginia Tech Advanced Research Computing's (ARC) Cascades cluster which utilizes standard nodes that contain 2 Intel Xeon E5-2683 v4 CPUs. 
The CP approximations were computed using the standard alternating least squares (ALS) method.\cite{VRG:kroonenberg:1980:P,VRG:beylkin:2002:PNAS} 
In this work, as was done in our previous work,\cite{VRG:pierce:2021:JCTC} initial CP3 factors were generated using quasi-random numbers taken from the uniform distribution on [-1,1].
As discussed earlier in this work, we utilized both optimized CP3 factors and quasi-random numbers as an initial guess strategy for the CP4 decomposition.

Assessment of the CP factorizations utilized the a small subset of 5 complexes from the S66 (S66/5) benchmark set of weakly-bound complexes\cite{VRG:rezac:2011:JCTC}\footnote{1 water \dots water, 2 water \dots \ce{MeOH}, 3 water \dots \ce{MeNH2}, 4 \ce{MeNH2} \dots \ce{MeOH}, 5 ethyne \dots water (CH-O)}. 
The S66 geometries were taken from the Benchmark Energy and Geometry Database (BEGDB).\cite{Rezac:2008:CCCC} 
Additionally, the  a conformer of (\ce{H2O})$_{12}$\cite{VRG:jorgensen:1983:JCP,Wales:1998:CPL} is used to demonstrate the capacity of the reduced-scaling CP4 ALS optimization strategy.
All of the above computations utilized the aug-cc-pVDZ\cite{Dunning:1989:JCP,VRG:kendall:1997:TCA} (aVDZ) orbital basis set (OBS). The 2-electron interaction tensors were approximated using standard Coulomb-metric density fitting using the aug-cc-pVDZ-RI (abbreviated as aVDZ-RI) density fitting basis set (DFBS).\cite{VRG:weigend:2002:JCP}
Assessment of the basis set variation in the performance of CP4 used the following additional OBS/DFBS pairs: the cc-pVDZ-F12 (abbreviated as DZ-F12)\cite{VRG:peterson:2008:JCP} OBS paired with the aVDZ-RI DFBS, the aug-cc-pVTZ (aVTZ) OBS\cite{Dunning:1989:JCP,VRG:kendall:1997:TCA} paired with the aug-cc-pVTZ-RI\cite{VRG:weigend:2002:JCP} (aVTZ-RI) DFBS, and  the cc-pVTZ-F12\cite{VRG:peterson:2008:JCP} (TZ-F12) OBS paired with the aVTZ-RI DFBS.

Lastly, to make comparisons between different systems the CP ranks will be defined in units of size $X$ of the DF fitting (``auxiliary'') basis, which grows proportionally to the number of atoms in the system.

\section{\label{sec:results}Results}

\begin{figure}[ht!]
\begin{subfigure}{0.45\textwidth}
        \includegraphics[width=\columnwidth]{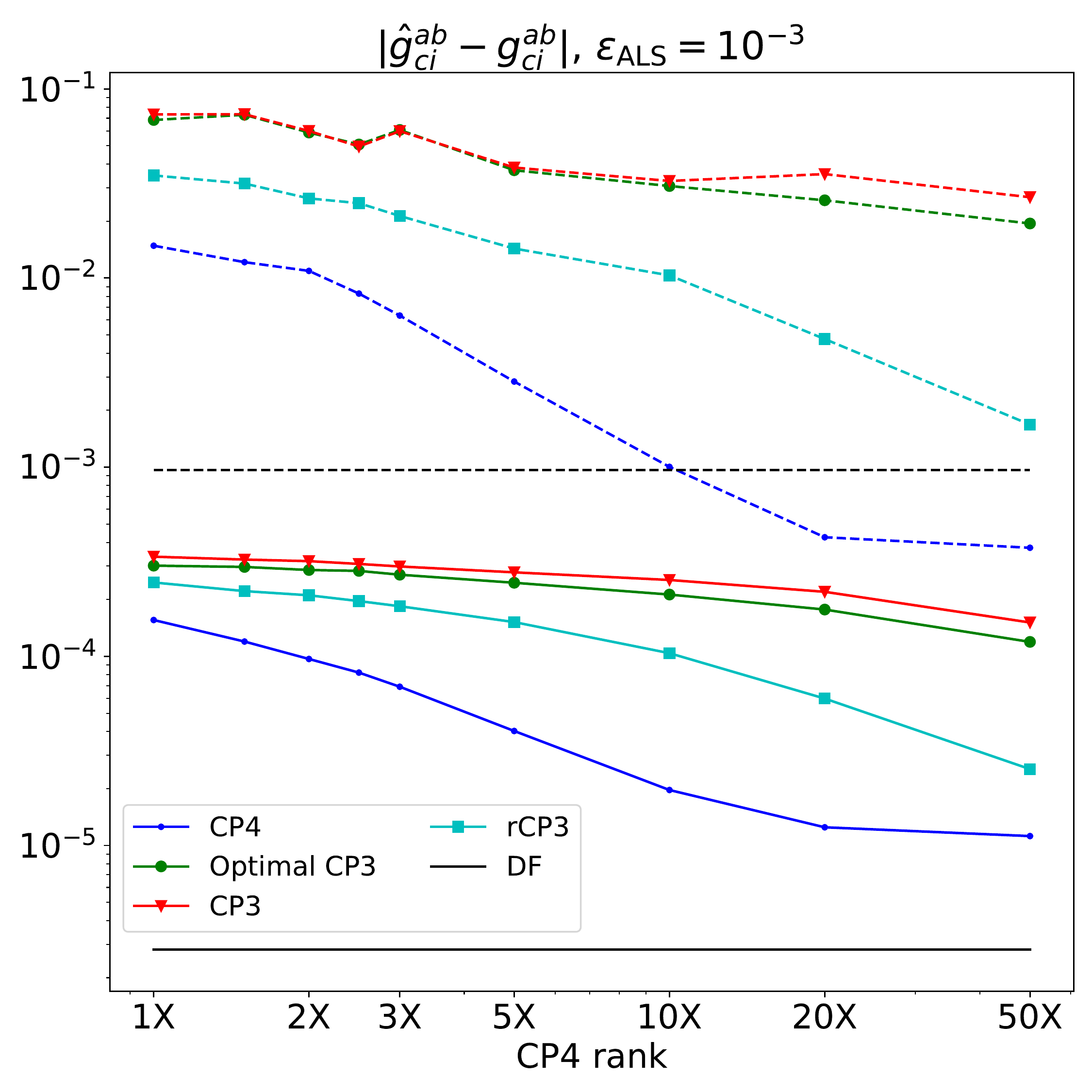}
        \caption{}
        \label{fig:Elabci_1e3}
\end{subfigure}
\begin{subfigure}{0.45\textwidth}
\includegraphics[width=\columnwidth]{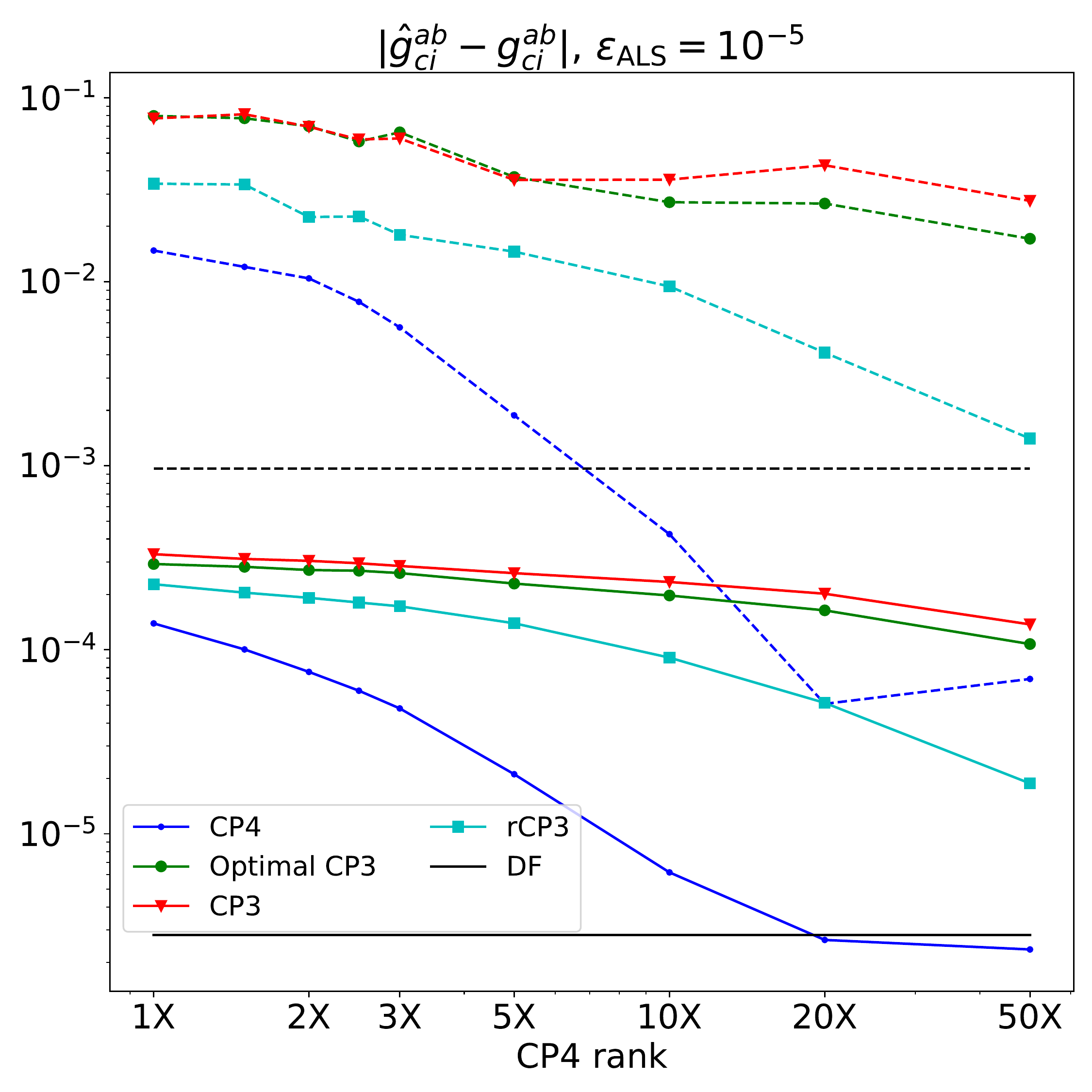}
    \caption{}
    \label{fig:Elabci_1e5}
\end{subfigure}
\caption{ Absolute average element error for approximated $g^{ab}_{ci}$ tensor using an ALS solver threshold\cite{VRG:pierce:2021:JCTC} of (a) $\epsilon_\mathrm{ALS}=10^{-3}$ and (b) $\epsilon_\mathrm{ALS}=10^{-5}$. Solid lines represents mean error and dashed lines represent max error. Note that the CP4 rank is in units of a factor times the DF fitting basis, a metric which grows linearly with rank. Data was collected using the S66/5 dataset with a aVDZ/aVDZ-RI basis}
\label{fig:ElErG_abci}
\end{figure}

\begin{figure}[ht!]
\begin{subfigure}{0.45\textwidth}
        \includegraphics[width=\columnwidth]{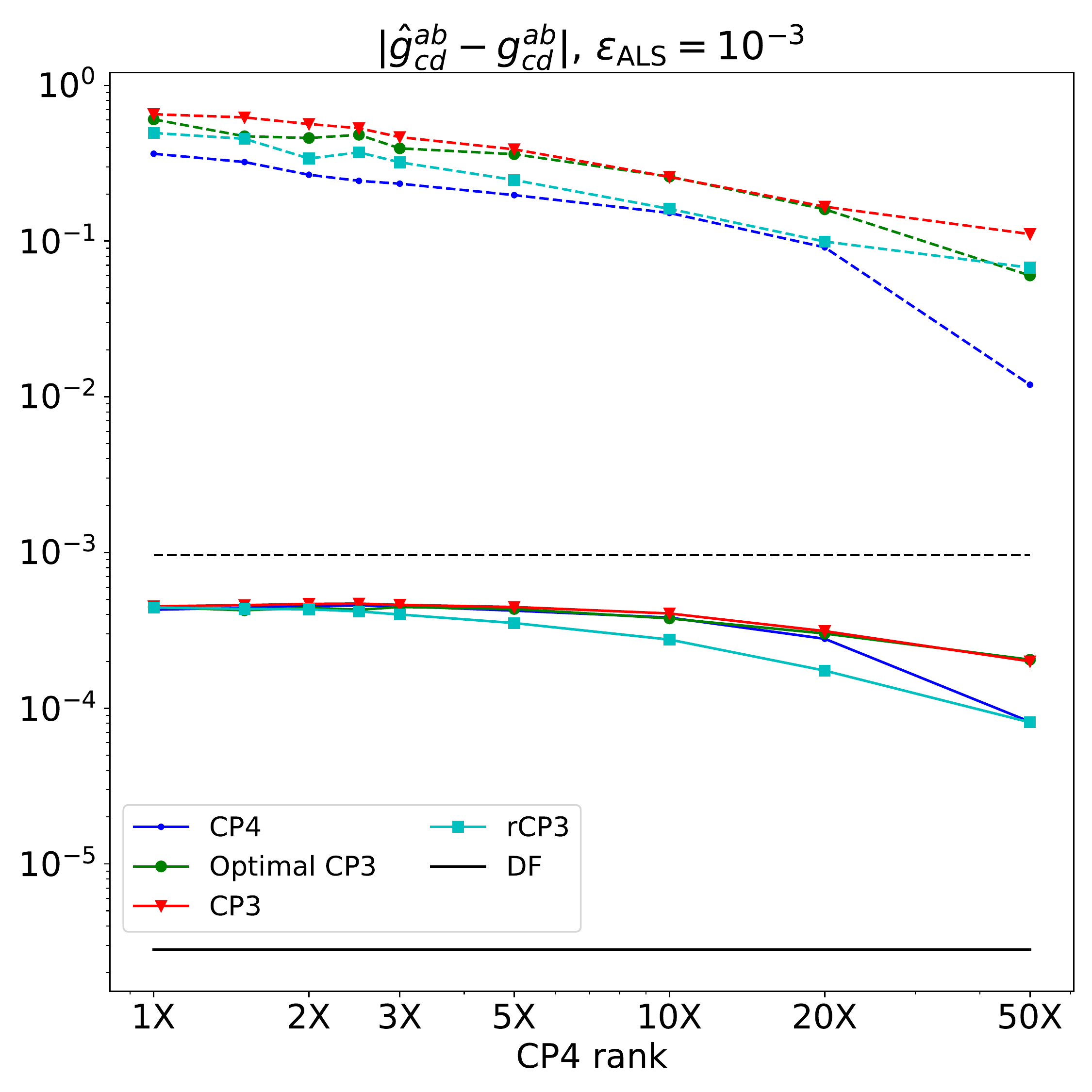}
        \caption{}
        \label{fig:Elabci}
\end{subfigure}
\begin{subfigure}{0.45\textwidth}
\includegraphics[width=\columnwidth]{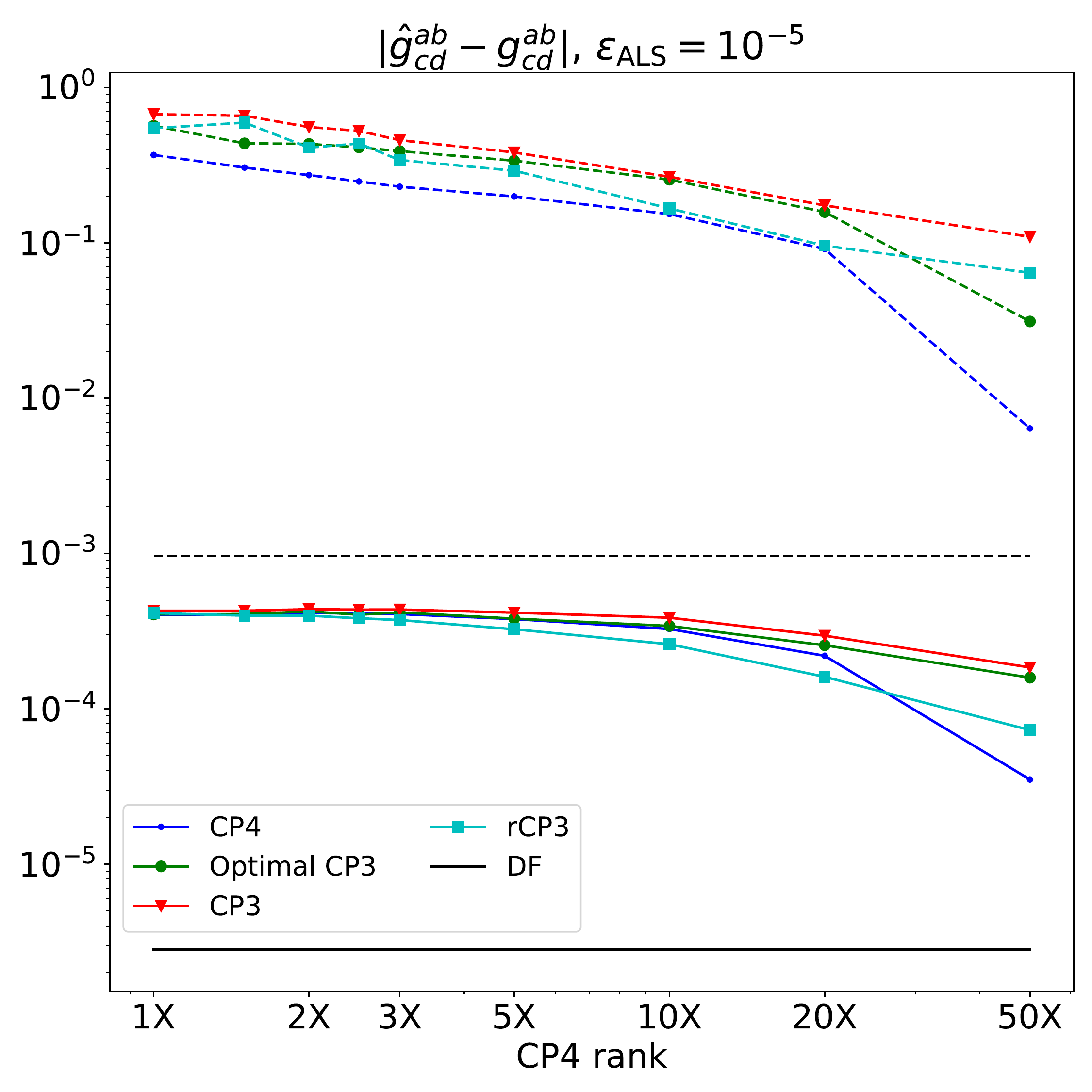}
    \caption{}
    \label{fig:Elabcd_1e5}
\end{subfigure}
\caption{Absolute average element error for approximated $g^{ab}_{cd}$ tensor using an ALS solver threshold\cite{VRG:pierce:2021:JCTC} of (a) $\epsilon_\mathrm{ALS}=10^{-3}$ and (b) $\epsilon_\mathrm{ALS}=10^{-5}$. Solid lines represents mean error and dashed lines represent max error. Note that the CP4 rank is in units of a factor times the DF fitting basis, a metric which grows linearly with rank. Data was collected using the S66/5 dataset with a aVDZ/aVDZ-RI basis}
\label{fig:ElErG_abcd}
\end{figure}

\cref{fig:ElErG_abci,fig:ElErG_abcd}~~illustrate the accuracy of the CP4 approximation compared to CP3-DF-based approximations. To compare tensor network approximations to each other it is necessary to map all of them to a common network topology, i.e., the CP3-DF tensor networks are mapped their CP4 counterpart via \cref{eq:g_THC_as_CP4}. Since a CP3-DF network with rank $R_\text{CP3}$ is represented exactly by a CP4 network with rank $R_\text{CP4} = R_\text{CP3}^2$ (see the discussion around \cref{eq:g_THC_as_CP4}), the CP4 network of rank $R_\text{CP4}$ are compared to the CP3 counterparts of rank $R_\text{CP3} = \sqrt{R_\text{CP4}}$. Furthermore, since the robust CP3-DF approximant (\cref{eq:g_rCP3DF}) is a sum of 3 contributions it makes sense to use even an lower CP3 rank ($\sim\sqrt{R_\text{CP4}/3}$) when comparing to a CP4 network. For the sake of simplicity, we conservatively use the same CP3 rank for all CP3-based networks.

In \cref{fig:ElErG_abci,fig:ElErG_abcd}~~the \{dashed, solid\} lines denoted ``DF'' show the \{maximum, average\} elementwise errors due to the DF approximation of the exact $g$ tensors, while the other lines show the corresponding elementwise errors associated with the CP decomposition of the DF-approximated $g$ tensors.
The lines labeled ``CP3'' correspond to the THC-like CP3-DF approximation of the $g$ tensors (\cref{eq:g_THC_1}) and the lines labeled ``rCP3'' correspond to the robust CP3-DF approximation ( \cref{eq:g_rCP3DF}).
The lines labeled ``CP4'' correspond to the CP4 decomposition using the DF-approximated $g$ tensor in the CP4 ALS optimization.
Finally, the lines labeled ``Optimal CP3'' correspond to the CP3-DF approximation (\cref{eq:g_THC_1}) in which the constituent  factors were optimized to make stationary an {\em augmented} CP loss function defined for, e.g., $g^{ab}_{ci}$ as 
\begin{align}
\label{eq:opt-cp3-f}
    f(\mathbf{x}_1, \mathbf{x}_2) &= \frac{1}{2}\| B^{aX}_{c} B^{bX}_{i} - \hat{B}^{aX}_{c} \hat{B}^{bX}_{i} \|^2 
    + \frac{1}{2}\| B^{aX}_{c} - \hat{B}^{aX}_{c}\|^2 
    + \frac{1}{2}\|B^{bX}_{i} - \hat{B}^{bX}_{i} \|^2,
\end{align}
where $\mathbf{x}_{\{1,2\}}$ are the CP3 parameters defining \{$\hat{B}^{aX}_{c}$ , $\hat{B}^{bX}_{i}$\}.
The augmented CP loss function is designed to simultaneouly solve 3 {\em coupled} least-squares problems, namely (1) finding the optimal CP3 factors for approximating the (DF-approximated) $g^{ab}_{ci}$ using the CP3-DF network, (2,3) finding the optimal CP3 factors for approximating the $\{ B^{aX}_{c}, B^{bX}_{i} \}$ DF factors, respectively. Coupling the three CP problems helps the ALS solver to avoid excessively large steps that may occur when solving problem 1 alone, which stems from lack of convergence guarantees for the ALS algorithm in general.\cite{VRG:kolda:2009:SR} Similar modifications of the objective functions to penalize large steps have been used by others.\cite{VRG:goings:2022:PNASU} The use of more robust direct minimization solvers for the CP and general tensor network optimization problems (e.g., nonlinear quasi-Newton solvers with step restriction) should alleviate the need for augmenting the CP3-DF objective function.


As \cref{fig:ElErG_abci,fig:ElErG_abcd} clearly demonstrate, at equivalent ranks the CP4 approximation is more accurate than the CP3-DF counterparts. CP4 even surpasses the accuracy of the robust CP3-DF variant which we recently showed to be more accurate than the corresponding PS-like and THC-like CP3-DF counterparts. Namely, CP4 is always more accurate than rCP3-DF for the $g^{ab}_{ci}$ tensor and surpasses accuracy of the latter with largest CP4 rank for the $g^{ab}_{cd}$ tensor.
However, notice that at modest ranks the accuracy of the CP4 approximation of $g^{ab}_{cd}$ was found to be lower than that of the robust CP3 approximation, see the solid lines in \cref{fig:ElErG_abcd}. As discussed above, this is likely due to the overly conservative choice of the relationship between the CP3 and CP4 ranks for the robust case (i.e. the rank for the rCP3 approximation should have been smaller by a factor of $\sqrt{3}$).

CP4 errors usually, but not always, decrease monotonically: for example, the average CP4 error for $g^{ab}_{ci}$ (\cref{fig:Elabci_1e3}) does not decrease substantially when the CP rank is increased from $20X$ to $50X$. This can be rationalized if we recall that the error of an approximate CP4 decomposition is due to 2 factors: finiteness of the CP4 rank (hence controlled by the rank) and the suboptimality of the factors (controlled by the ALS convergence threshold $\epsilon_{\rm ALS}$). 
As the CP4 rank increases the factor optimization error becomes dominant. Indeed, as the ALS threshold decreases from the $10^{-3}$ to $10^{-5}$ this artifact disappears as well as the total errors for large ranks ($\geq5X$) decrease significantly. 
This result is consistent with our previous finding for the CP3 optimization.\cite{VRG:pierce:2021:JCTC}
It is well understood that 
more robust CP solvers\cite{VRG:acar:2011:JC,VRG:espig:2012:NM,VRG:kazeev:2010:CMaMP,VRG:phan:2013:SJMA&A} can be employed to improve the convergence rate related issues.
Since our focus here is primarily on addressing the universal concerns (the cost of gradient evaluation, initial guess) that apply to all CP solvers, here we only use ALS and postpone the use of improved CP solvers to later studies.

\begin{figure}[!b]
    \begin{subfigure}{0.5\textwidth}
        \includegraphics[width=\columnwidth]{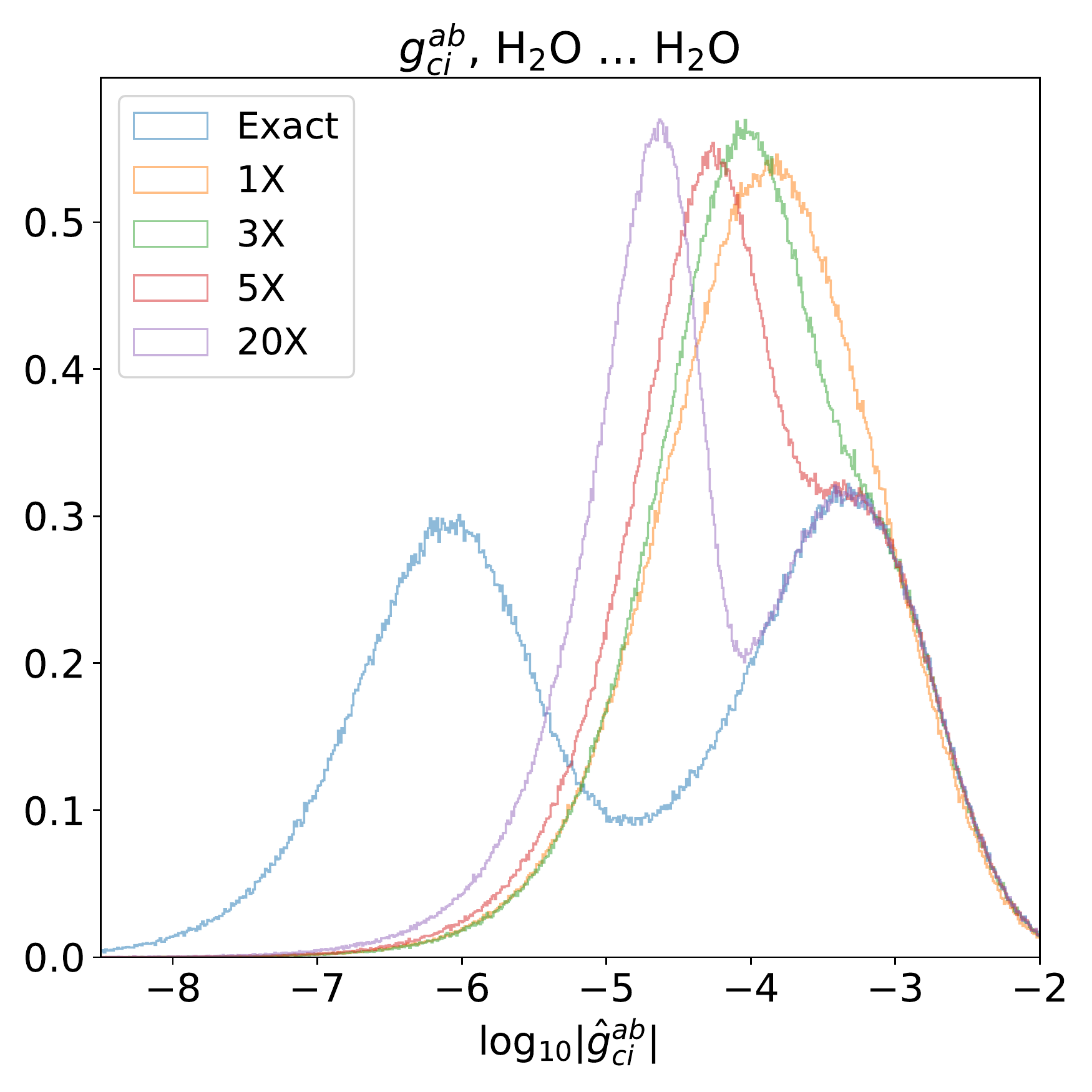}
        \caption{}
        \label{fig:g_abci_13}
    \end{subfigure}\hfill
    \begin{subfigure}{0.5\textwidth}
        \includegraphics[width=\linewidth]{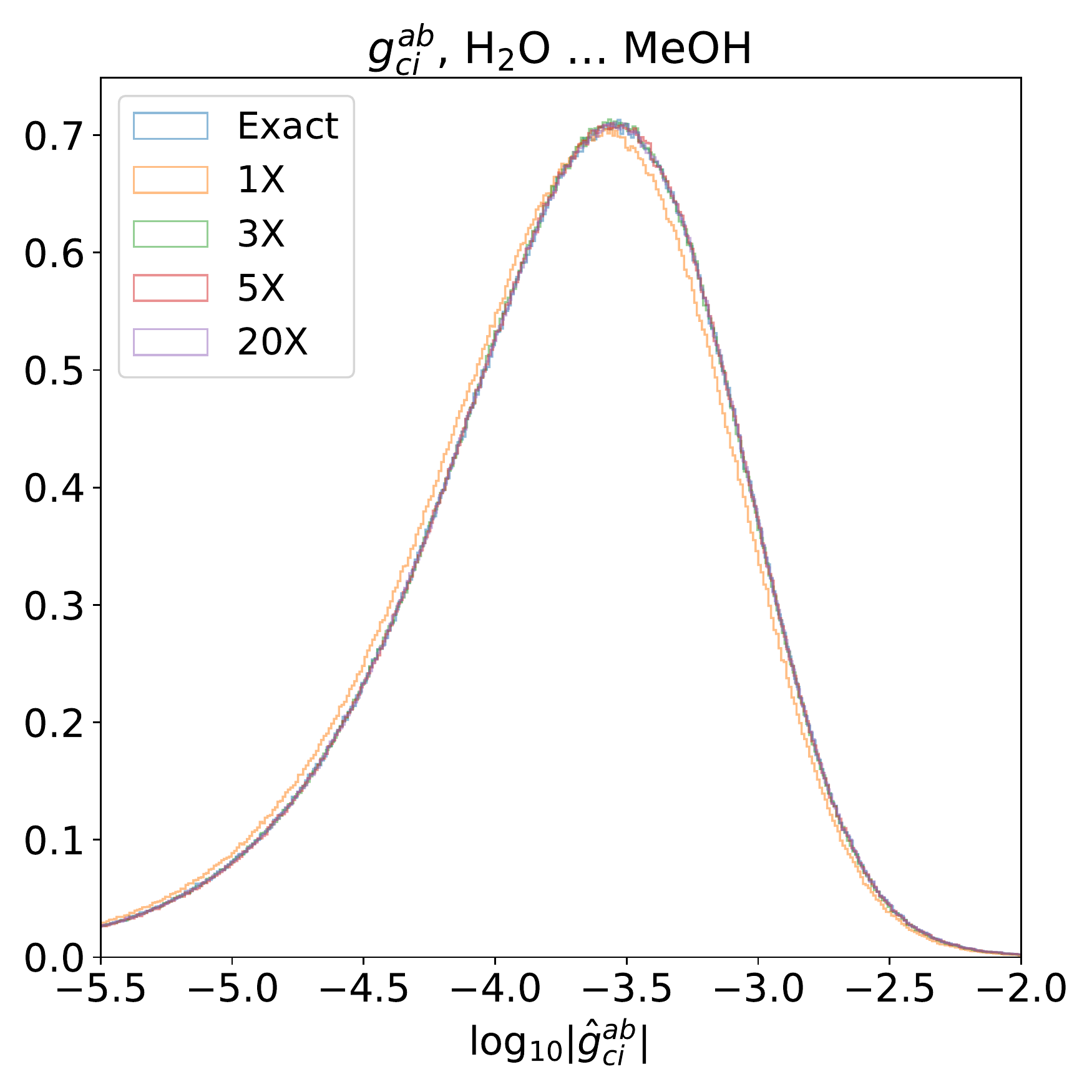}
        \caption{}
        \label{fig:g_abci_14}
    \end{subfigure}\hfill
    \begin{subfigure}{0.5\textwidth}
        \includegraphics[width=\columnwidth]{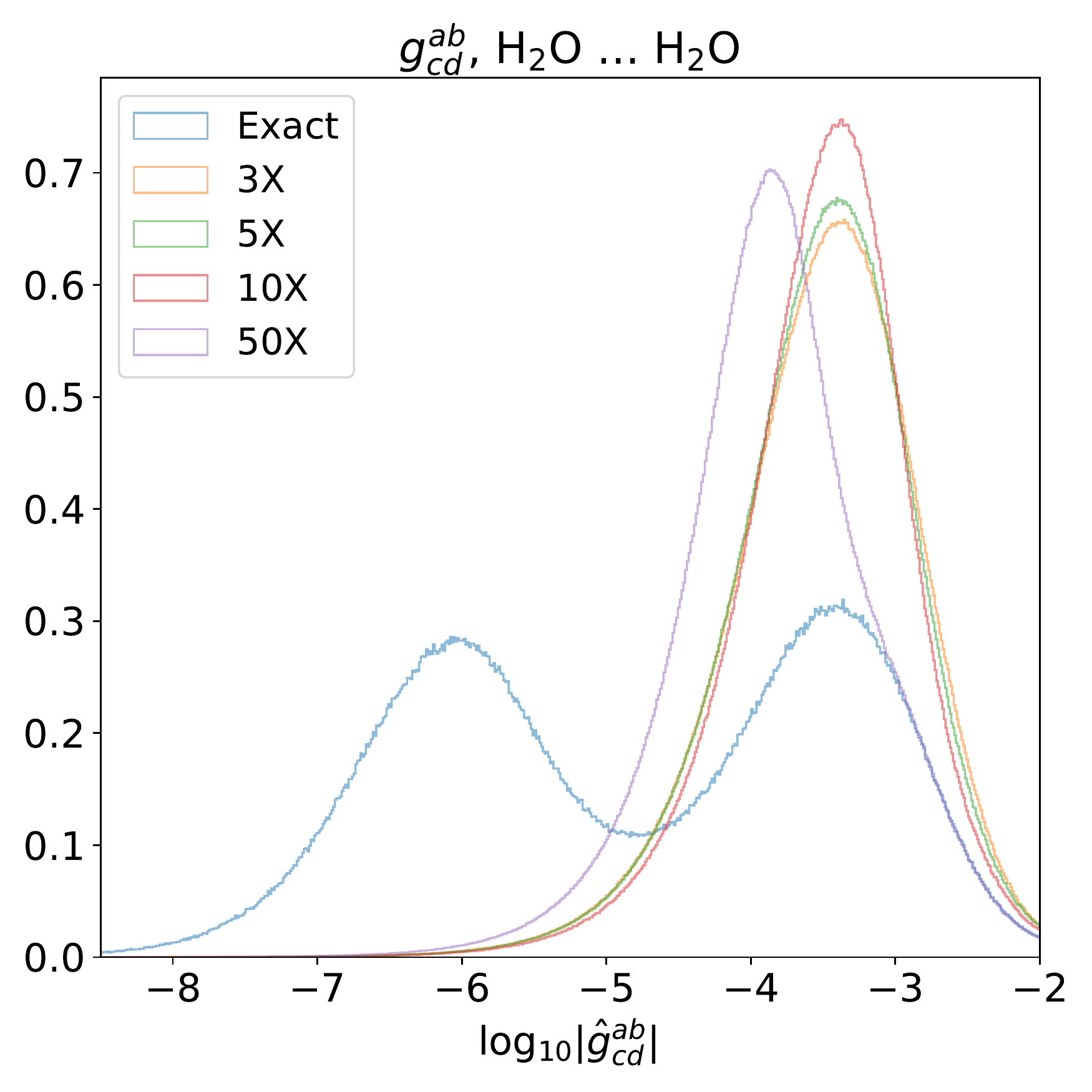}
        \caption{}
        \label{fig:g_abcd_13}
    \end{subfigure}\hfill
    \begin{subfigure}{0.5\textwidth}
        \includegraphics[width=\linewidth]{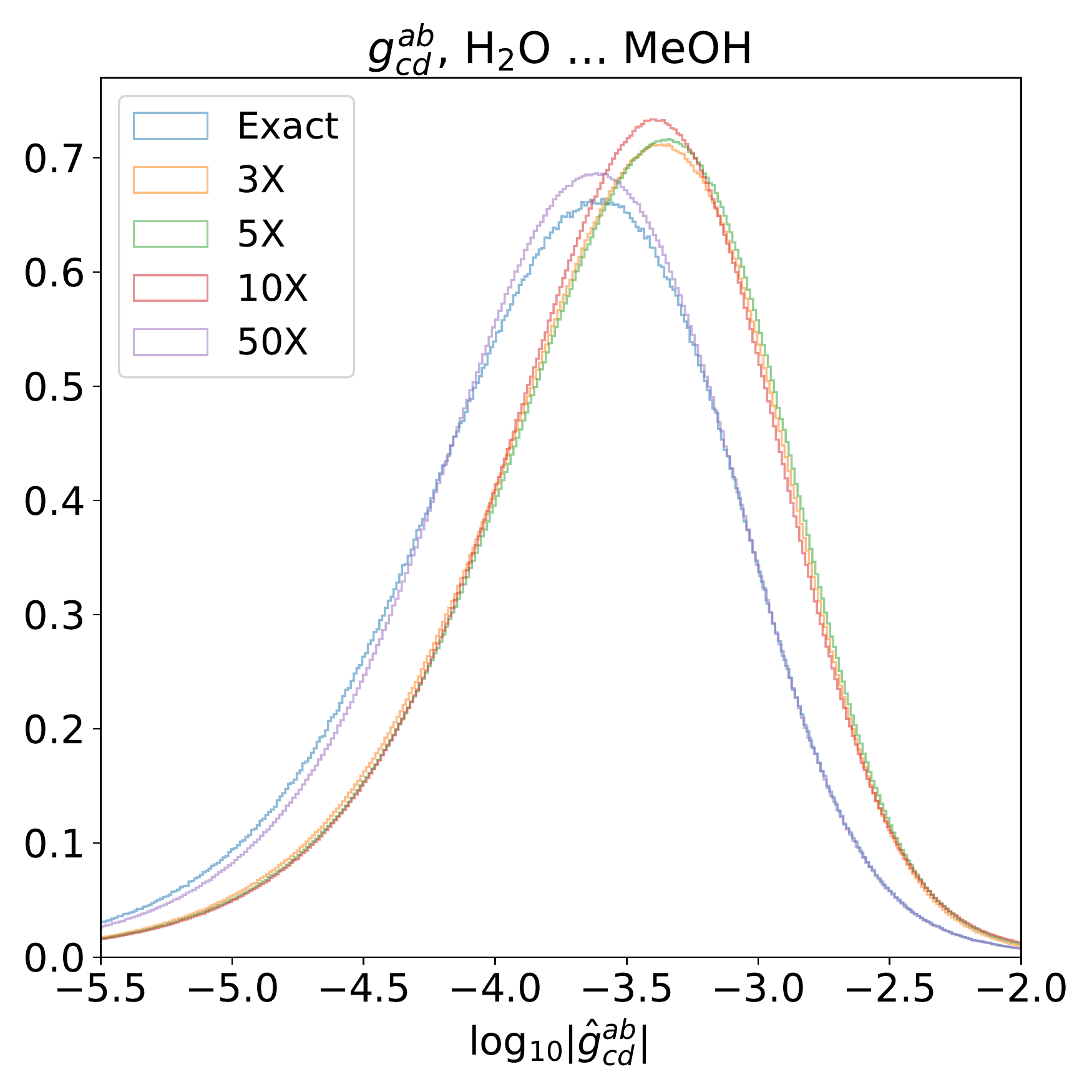}
        \caption{}
        \label{fig:g_abcd_14}
    \end{subfigure}\hfill
\caption{Distributions of the absolute element values for $g^{ab}_{ci}$ [subfigures (a) and (b)] and $g^{ab}_{cd}$ [subfigures (c) and (d)] tensors for representative systems. ``Exact'' refers to the DF-based tensor free of CP approximations, while ``nX'' corresponds to the counterpart approximated with CP4 decomposition of rank $nX$. All
computations utilized a aVDZ/aVDZ-RI basis and $\epsilon_\mathrm{ALS} = 10^{-3}$.}
\label{fig:hist_g}
\end{figure}

Another important phenomenon becomes apparent when we plot the distribution of tensor element magnitudes for the two types of tensors in the two test systems (\cref{fig:hist_g}).
One of the two test systems (namely, the water dimer) has geometrical symmetry that makes many tensor elements nearly zero; note that our orbital solver does not exploit geometrical symmetry and therefore the zeroes are {\em soft} (i.e., the elements that are zero by symmetry are merely small).
The presence of symmetry results in markedly bimodal distributions of the element magnitudes for both types of tensors for the water dimer (\cref{fig:g_abci_13,fig:g_abcd_13}),
with the ``left'' peak corresponds to the elements of the exact tensor that are nearly zero by symmetry.
However, the distribution of the elements in the CP-approximated tensors are unimodal, except for the highest-rank approximants to the $g^{ab}_{ci}$ tensor.
As the CP4 rank increases the largest tensor elements get approximated better and better, whereas the near-zero elements are not well-approximated (in relative sense) even with the largest CP4 ranks employed here.
The inability to preserve the blocked structure of the tensor is due to the global character of the CP4 factorization; to impose a blocked structure on the CP approximant would sacrifice the simplicity of the ansatz, hence was not done here.

Besides the block-sparse structure of the tensor approximate CP decomposition also fails to preserve permutational symmetries of the tensor. For example, consider permutational symmetry
of the (real) $g^{ab}_{cd}$ tensor:
\begin{align}
\label{eq:g-abcd-8fold-symmetry}
g^{ab}_{cd} = g^{ad}_{cb} = g^{cb}_{ad} = g^{cd}_{ab} = g^{ba}_{dc} = g^{bc}_{da} = g^{da}_{bc} = g^{dc}_{ba}.
\end{align}
The CP4 approximant $\hat{g}^{ab}_{cd}$ (\cref{eq:g_CP4})
in general (i.e., for an arbitrary choice of CP4 parameters) lacks the symmetries of the exact tensor.
Since it is possible in general to represent $g^{ab}_{cd}$ exactly in the CP4 form, the CP4 factors must have particular structure for the approximant to possess the exact symmetries of the tensor. Hence we inspected the magnitude of the deviation of the CP4-approximated $g^{ab}_{cd}$ tensors from the exact 8-fold symmetry in \cref{eq:g-abcd-8fold-symmetry} as a function of the CP4 rank. Table \cref{tab:symm} illustrates how the error of the standard non-symmetrized
CP4 approximant compares to that of a (posteriori) symmetrized CP4 approximant,
\begin{align}
\label{eq:g-abcd-S-CP4}
\hat{S} \hat{g}^{ab}_{cd} \equiv \frac{1}{8}\left( \hat{g}^{ab}_{cd} + \hat{g}^{ad}_{cb} + \hat{g}^{cb}_{ad} + \hat{g}^{cd}_{ab} + \hat{g}^{ba}_{dc} + \hat{g}^{bc}_{da} + \hat{g}^{da}_{bc} + \hat{g}^{dc}_{ba} \right).
\end{align}
There are several conclusion that can be drawn from this data. First, the violation of the exact symmetries is relatively small compared to the error due to the error of the CP4 approximant, ranging from $\sim1\%$ with $R_\mathrm{CP4}=X$ to $\sim 30\%$ with $R_\mathrm{CP4}=50X$. Second, enforcing the symmetry always reduces the error. Since for some applications it may be worthwhile to enforce the symmetry by construction it would be potentially useful in the future to explore symmetry-adapted CP ansatze.

\begin{table}[!ht]
        \label{tab:symm}
        \centering
        \begin{tabular}{| l | l | ccccccccc |}
        \hhline{===========}
            System & Measure & \multicolumn{9}{c|}{$R_\mathrm{CP4}/X$} \\
                         &                & 1 & 1.5 & 2 & 2.5 & 3  & 5 & 10 & 20 & 50\\
            \hline
& $\|g - \hat{g} \|$ & 17.0 & 16.1 & 15.3 & 14.6 & 13.9 & 11.6 & 8.01 & 3.15 & 0.854 \\
(\ce{H2O})$_{2}$ & $\|g - \hat{g}_\mathrm{sym} \|$ & 16.9 & 15.8 & 14.9 & 14.2 & 13.5 & 11.2 & 7.61 & 2.63 & 0.612 \\
& $\|\hat{g} - \hat{g}_\mathrm{sym} \|$ & 2.34 & 2.96 & 3.38 & 3.45 & 3.29 & 2.77 & 2.52 & 1.74 & 0.595 \\
\hline
& $\|g - \hat{g} \|$ & 23.1 & 22.1 & 21.3 & 20.5 & 19.9 & 17.6 & 13.9 & 8.71 & 1.05 \\
\ce{H2O} ... \ce{MeOH} & $\|g - \hat{g}_\mathrm{sym} \|$ & 22.9 & 21.8 & 20.9 & 20.1 & 19.4 & 17.1 & 13.3 & 8.01 & 0.763 \\
& $\|\hat{g} - \hat{g}_\mathrm{sym} \|$ & 3.31 & 3.60 & 4.01 & 4.13 & 4.12 & 3.84 & 3.88 & 3.41 & 0.718 \\
\hhline{===========}
        \end{tabular}
        \caption{Comparison of standard (\cref{eq:g_CP4}) and symmetrized (\cref{eq:g-abcd-S-CP4}) CP4 approximants to DF-approximated $g^{ab}_{cd}$ evaluated with the aVDZ/aVDZ-RI basis set pair; the corresponding error 2-norms are denoted by $\|g - \hat{g} \|$ and $\|g - \hat{g}_\mathrm{sym} \|$, respectively. $\|\hat{g} - \hat{g}_\mathrm{sym} \|$ denotes the 2-norm of the difference between the standard and symmetrized CP4 approximants.}
    \end{table}

\begin{figure}[htb]
    \begin{subfigure}{0.5\textwidth}
        \includegraphics[width=\columnwidth]{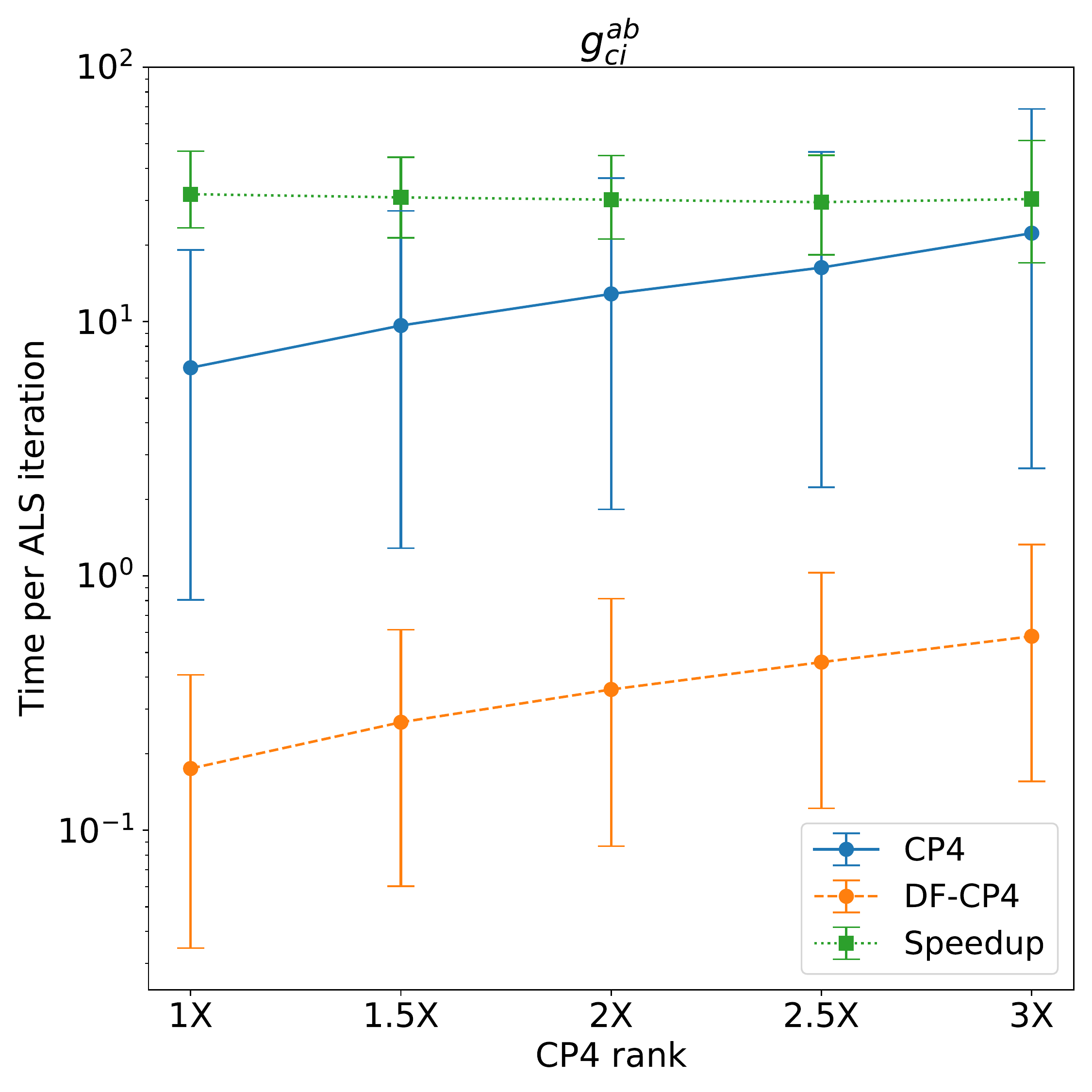}
        \caption{}
        \label{fig:abci-iter}
    \end{subfigure}\hfill
    \begin{subfigure}{0.5\textwidth}
        \includegraphics[width=\linewidth]{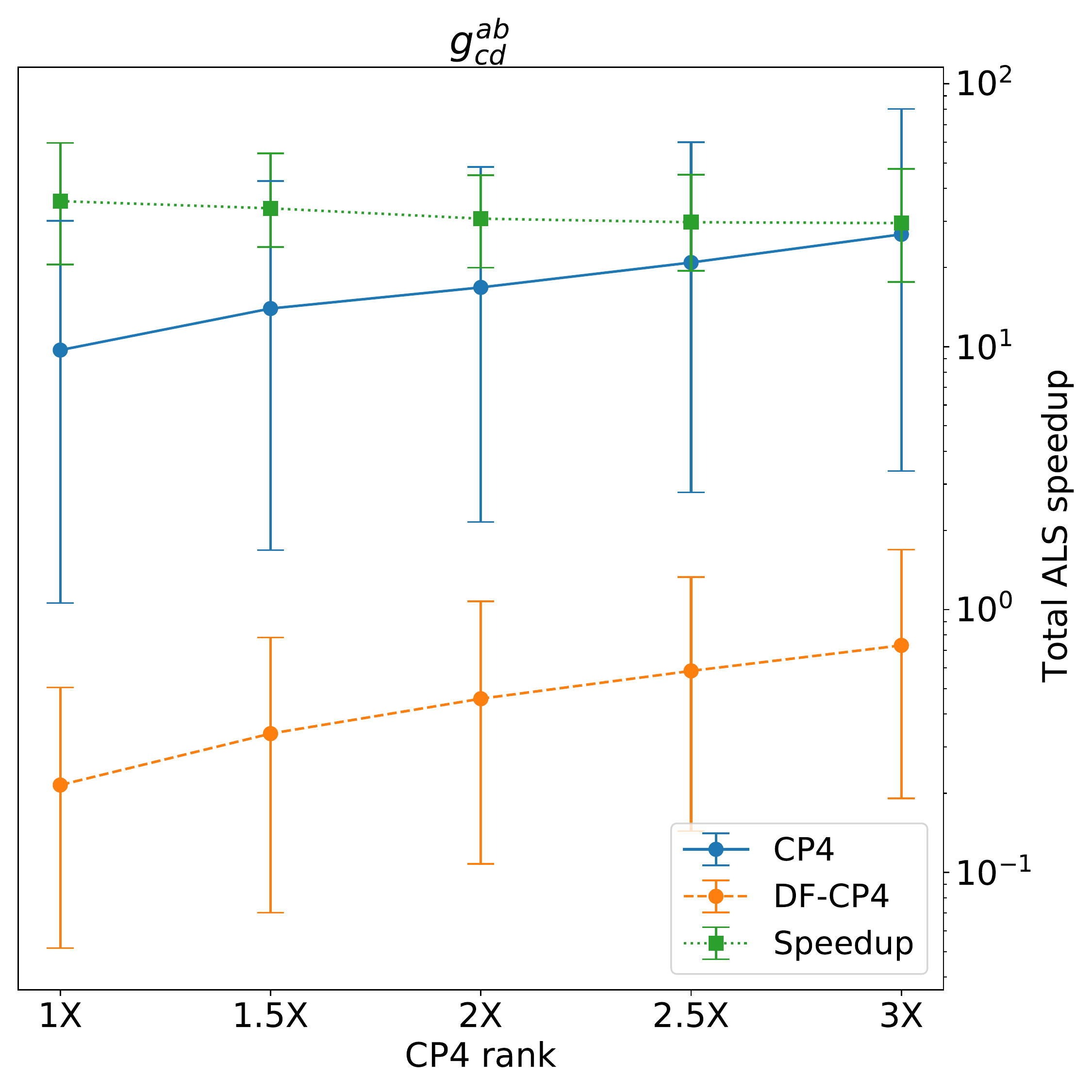}
        \caption{}
        \label{fig:abcd-iter}
    \end{subfigure}\hfill
\caption{The cost per iteration of the exact and DF approximated CP4 ALS decomposition and the total speedup associated with the DF approximation, i.e. $\frac{t_\mathrm{exact}}{t_\mathrm{DF}}$. The error bars represent the maximum and minimum per iteration cost. Data was collected using the S66/5 dataset and the aVDZ/aVDZ-RI basis pair.}
\label{fig:perIter}
\end{figure}

\begin{figure}[!b]
    \begin{subfigure}{0.5\textwidth}
        \includegraphics[width=\columnwidth]{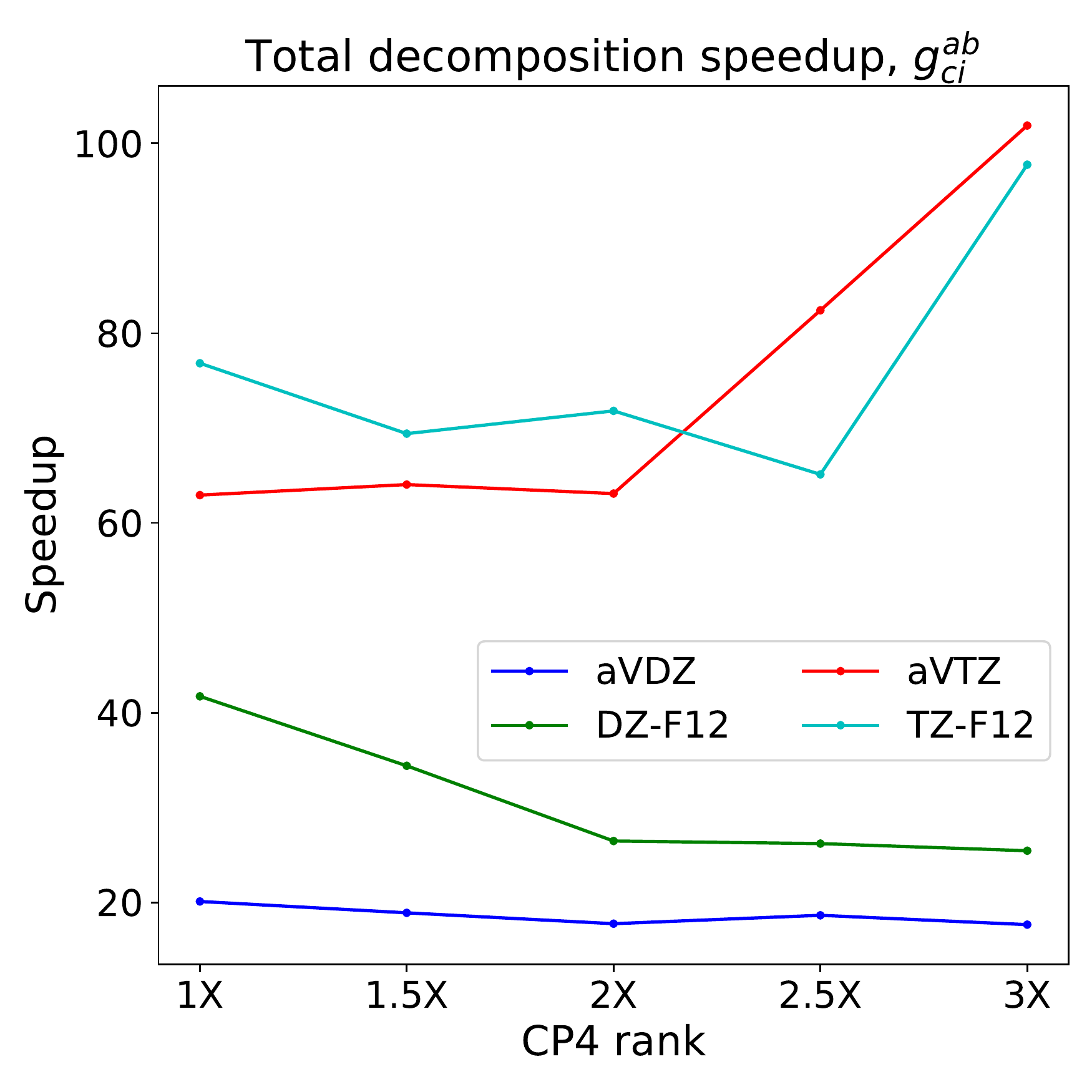}
        \caption{}
        \label{fig:abci-totSpeed}
    \end{subfigure}\hfill
    \begin{subfigure}{0.5\textwidth}
        \includegraphics[width=\linewidth]{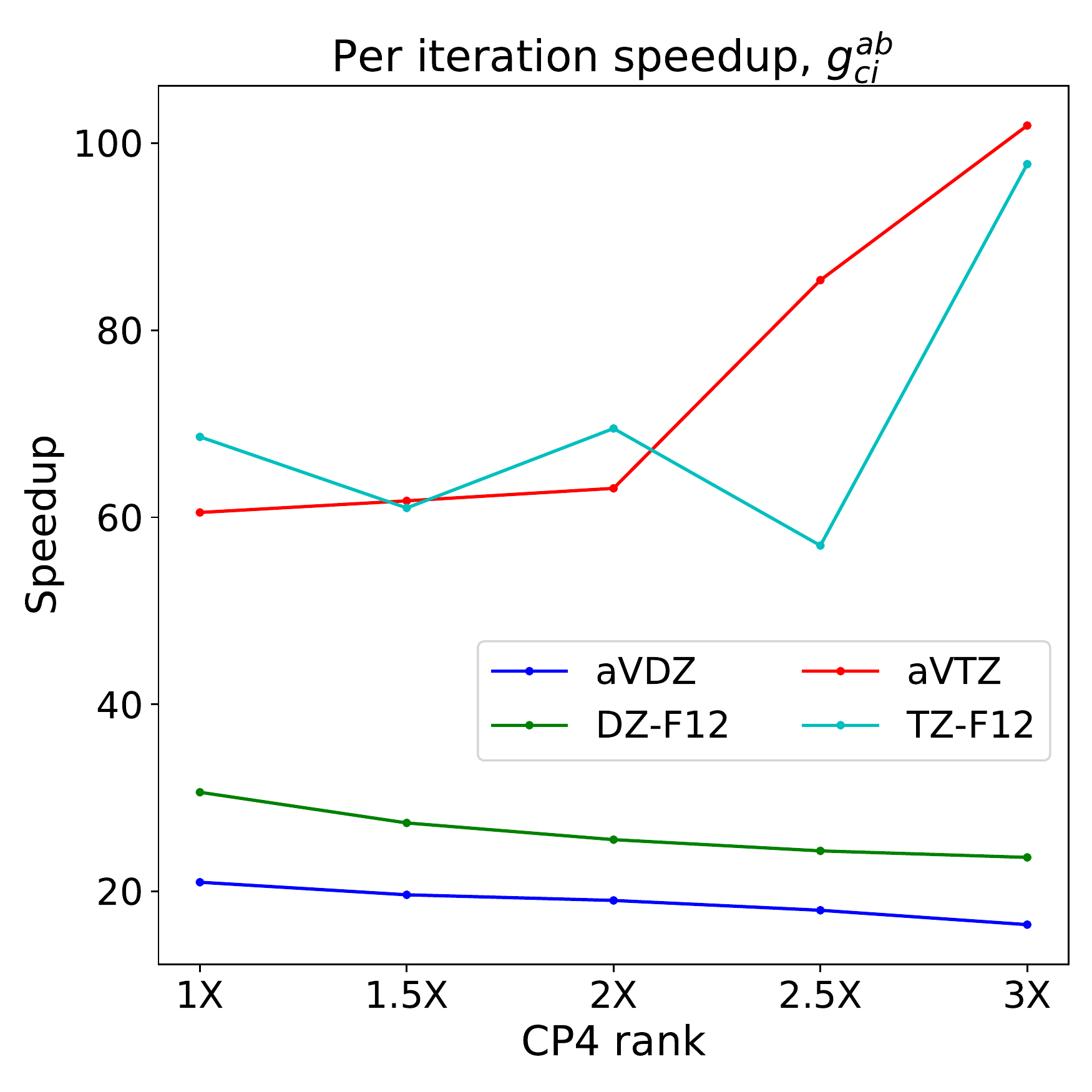}
        \caption{}
        \label{fig:abcd-totSpeed}
    \end{subfigure}\hfill
    \begin{subfigure}{0.5\textwidth}
        \includegraphics[width=\columnwidth]{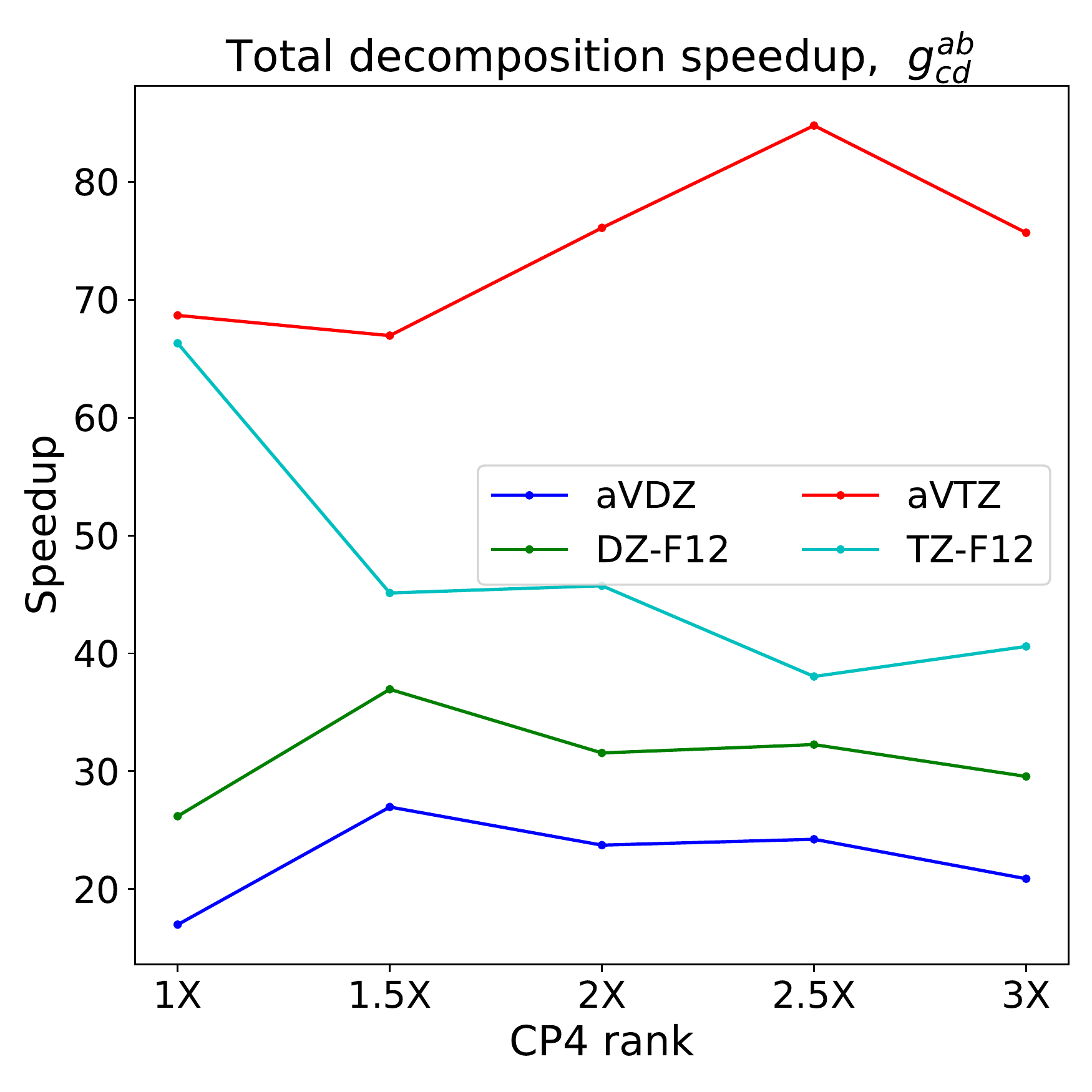}
        \caption{}
        \label{fig:abci-perIter}
    \end{subfigure}\hfill
    \begin{subfigure}{0.5\textwidth}
        \includegraphics[width=\linewidth]{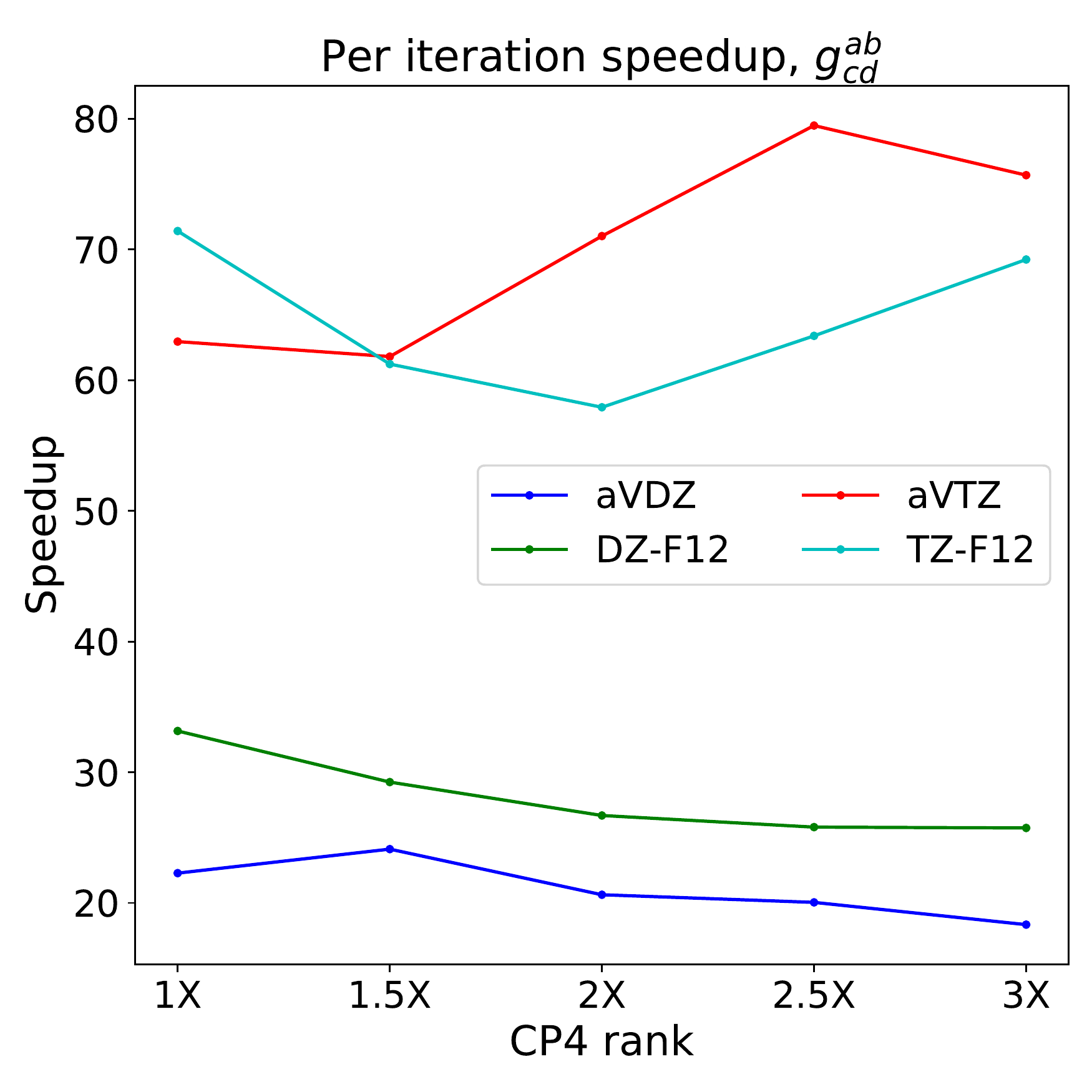}
        \caption{}
        \label{fig:abcd-perIter}
    \end{subfigure}\hfill
\caption{The total speedup of the CP4 ALS optimization of the (a) $g^{ab}_{ci}$ tensor 
and (c) $g^{ab}_{cd}$ tensor and the per iteration speedup of the 
(b) $g^{ab}_{ci}$ tensor and
(d) $g^{ab}_{cd}$ tensor using a (\ce{H2O})$_{2}$ cluster from the S66 dataset and 4 different orbital and DF basis sets}
\label{fig:basis_set_test}
\end{figure}

\cref{fig:perIter} illustrates the wall time per ALS iteration with and without the DF approximation of the the CP4 gradient (\cref{eq:fCP-grad-SQ}) as well as the corresponding speedup (i.e., reduction in computational time per ALS iteration), versus the CP4 rank for the molecules in the S66/5 dataset. With the aVDZ/aVDZ-RI basis set pair the average observed reduction in per-iteration cost is $\sim30$.
The speedup figure is clearly basis-dependent, as illustrated using 4 different OBSs for the water dimer in \cref{fig:basis_set_test}; speedups as large as $100$ have been observed.

Finally, we use the DF-accelerated CP4 ALS optimization to decompose the $g^{ab}_{ci}$ tensor for a (\ce{H2O})$_{12}$ cluster in the aVDZ/aVDZ-RI basis and report the results in \cref{table:h12}.
The size of this tensor is 31 GB, which is two orders of magnitude larger than the largest tensors that have previously been decomposed using CP4 in the chemistry context.\cite{VRG:benedikt:2011:JCP}
Not only are we able to construct a CP4 decomposition of this high-cost integral tensor, but we can systematically improve the accuracy by increasing rank while maintaining a relatively small number of ALS iterations and thus low computational cost.
\begin{table}[!ht]
        \label{table:h12}
        \centering
        \begin{tabular}{|c|ccccc|}
        \hhline{======}
            CP4 Rank & 1.0 & 1.5 & 2.0 & 2.5 & 3.0 \\
            \hline
CP3 iterations     & 10/8  & 10/8 & 9/9  & 9/10 & 8/11 \\
CP4 iterations     & 21    & 21   & 22   & 24   & 24 \\
Per iteration cost (s)       & 9.62   & 15.5  & 22.6 & 30.8 & 29.8 \\
CP4 time (s)       & 202   & 325  & 498 & 739 & 716 \\
Max error  & $8.05 \times 10^{-3}$ & $7.72 \times 10^{-3}$ &$6.71 \times 10^{-3}$ & $6.17 \times 10^{-3}$ & $5.92 \times 10^{-3}$ \\ 
mean error  & $2.48 \times 10^{-5}$  &$2.06 \times 10^{-5}$  & $1.82 \times 10^{-5}$& $1.62 \times 10^{-5}$ & $1.49 \times 10^{-5}$\\ 
\hhline{======}
        \end{tabular}
        \caption{Computation of the CP4 decomposition of the $g^{ab}_{ci}$ tensor for a (\ce{H2O})$_{12}$ cluster using DF approximated CP4 ALS optimization strategy and using CP3 factor matrices as 
        an initial guess using $R_\mathrm{CP3} = R_\mathrm{CP4}$.
        The CP3 iteration's column shows the number of iterations required to compute the CP3 decomposition of $B^{aX}_{c}/B^{c}_{iX}$.
        The CP4 time reported does not include the time it takes to construct the CP3 initial guess because there is no one best, unique way to construct the guess.
        The max and mean error are the element-wise error $|g^{ab}_{ci} - \hat{g}^{ab}_{ci}|$
        }
    \end{table}

\section{Summary and Perspective} \label{sec:summary}

We demonstrated how to efficiently compute full (4-way) CP decomposition of large 2-body interaction tensors of relevance to electronic structure simulations by (1) reducing the cost of the CP gradient evaluation via a generalized square root (SQ) decomposition (such as density fitting or Cholesky decomposition), and (2) using 3-way CP decomposition of the SQ factors to construct initial guess for the 4-way CP decomposition. These improvements permit us to compute CP decompositions with greater speed and lower storage requirements than previously possible. 4-way CP decomposition of Coulomb integral tensors with 3 and 4 virtual indices was demonstrated to be more compact and/or accurate than the corresponding combinations of 3-way CP decompositions (similar to the pseudospectral and THC approximations) for several small clusters in the standard S66 benchmark set; practicality of a DF-accelerated 4-way CP solver was demonstrated for the 3-virtual-index integral tensor of a system with 36 atoms.

The techniques described in this paper are more generally applicable than just for approximating the 2-body interaction tensors in the electronic structure domain. The key lesson here is that it can be much more efficient to compute a CP approximation of a target tensor using a representative tensor network in place of the (reconstructed or exact) tensor itself. This suggests that the CP factorization, which offers the most complete decoupling of the tensor modes compared to other tensor network topology, may be more broadly deployable than currently thought, by combining it with other tensor network approximations.  We will demonstrate productive uses of CP4 approximation in the context of correlated electronic structure models elsewhere soon.

\begin{acknowledgement}
This work was supported by the U.S. Department of Energy award DE-SC0022263 provided via the Scientific Discovery through Advanced Computing (SciDAC) program by the Offices of Advanced Scientific Computing Research (ASCR) and Basic Energy Sciences (BES). We also acknowledge Advanced Research Computing at Virginia Tech (www.arc.vt.edu) for providing computational resources and technical support that have contributed to the results reported within this paper.
\end{acknowledgement}

\bibliography{vrgrefs.bib,kmprefs.bib}

\end{document}